\documentclass[aps,prl,twocolumn,amsmath,amssymb,nofootinbib,superscriptaddress,floatfix,reprint,longbibliography]{revtex4-1}
\usepackage[dvips]{graphicx}
\usepackage{latexsym}
\usepackage{amsmath}
\usepackage{amsfonts}
\usepackage{amssymb}
\usepackage{bm}
\usepackage{color}
\usepackage{txfonts}
\usepackage{float}
\usepackage{url}
\usepackage[colorlinks=true, urlcolor=blue, linkcolor=blue, citecolor=blue]{hyperref}
\usepackage{ulem}
\begin{document}
	\newcommand{\fig}[2]{\includegraphics[width=#1]{#2}}
	\newcommand{\la}{{\langle}}
	\newcommand{\ra}{{\rangle}}
	\newcommand{\dg}{{\dagger}}
	\newcommand{\upa}{{\uparrow}}
	\newcommand{\dna}{{\downarrow}}
	\newcommand{\ab}{{\alpha\beta}}
	\newcommand{\ias}{{i\alpha\sigma}}
	\newcommand{\ibs}{{i\beta\sigma}}
	\newcommand{\hH}{\hat{H}}
	\newcommand{\hn}{\hat{n}}
	\newcommand{\hc}{{\hat{\chi}}}
	\newcommand{\hU}{{\hat{U}}}
	\newcommand{\hV}{{\hat{V}}}
	\newcommand{\br}{{\bf r}}
	\newcommand{\bk}{{{\bf k}}}
	\newcommand{\bq}{{{\bf q}}}
	\def\gsim{~\rlap{$>$}{\lower 1.0ex\hbox{$\sim$}}}
	\setlength{\unitlength}{1mm}
	\newcommand{{\vhf}}{$\chi^\text{v}_f$}
	\newcommand{{\vhd}}{$\chi^\text{v}_d$}
	\newcommand{{\vpd}}{$\Delta^\text{v}_d$}
	\newcommand{{\ved}}{$\epsilon^\text{v}_d$}
	\newcommand{{\vved}}{$\varepsilon^\text{v}_d$}
	\newcommand{{\tr}}{{\rm tr}}
	\newcommand{\pprl}{Phys. Rev. Lett. \ }
	\newcommand{\pprb}{Phys. Rev. {B}}

\title {The destiny of obstructed atomic insulator under correlation}
\author{Kun Jiang}
\email{jiangkun@iphy.ac.cn}
\affiliation{Beijing National Laboratory for Condensed Matter Physics and Institute of Physics,
	Chinese Academy of Sciences, Beijing 100190, China}
\affiliation{School of Physical Sciences, University of Chinese Academy of Sciences, Beijing 100190, China}

\author{Hongming Weng}
\email{hmweng@iphy.ac.cn}
\affiliation{Beijing National Laboratory for Condensed Matter Physics and Institute of Physics,
	Chinese Academy of Sciences, Beijing 100190, China}

\author{Jiangping Hu}
\email{jphu@iphy.ac.cn}
\affiliation{Beijing National Laboratory for Condensed Matter Physics and Institute of Physics,
	Chinese Academy of Sciences, Beijing 100190, China}
\affiliation{Kavli Institute of Theoretical Sciences, University of Chinese Academy of Sciences,
	Beijing, 100190, China}

\date{\today}

\begin{abstract}
The obstructed atomic insulators are insulators with both atomic limits and boundary states. In this work, we study the obstructed atomic insulators under correlation. We use the symmetry indicators by constructing many-body wavefunctions in momentum space to prove the obstruction properties in different models including  the SSH chain, anisotropic square lattice model, the quadrupole insulator model and etc. We demonstrate that the obstruction properties with boundary modes persist at large $U$ where the charge freedom is well-gapped, namely, this insulator phase can smoothly connect  to its Mott phase without Mott transition.   
\end{abstract}
\maketitle

\begin{figure}
	\begin{center}
		\fig{3.4in}{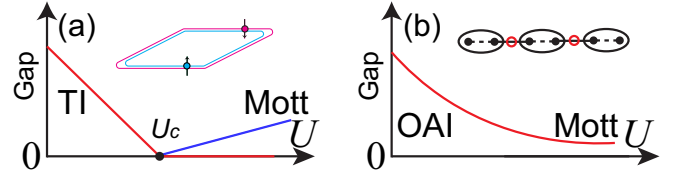}\caption{(a) The schematic phase diagram of TI under correlation U. The topological gap continuously decreases when increasing correlation strength U and vanishes at the phase transition point $U_c$. The Mott phase can be gapless magnetic order (red line) or other gapped systems (blue line) depending on model details. The inset illustrates a quantum spin hall insulator.
  (b) The schematic phase diagram of OAI under correlation U. OAI will evolve into the Mott phase without gap closing. The inset illustrates the OAI, where black sites are atomic sites and red circles are the Wannier centers.
			\label{fig0}}
	\end{center}
	\vskip-0.5cm
\end{figure}

\textit{Introduction}
How to find and classify phases of matter is one central theme in condensed matter physics. The discovery of topological insulators (TIs) and other topological phases profoundly changed our view on matter classification \cite{qi_RevModPhys.83.1057,kane_RevModPhys.82.3045}. 
Re-tracking the development of topological physics, the combination with density functional theory greatly boosts the development of predicting and identifying topological properties \cite{fangchen,xiangang_wan,Topological_quantum_chemistry,bernevig,zhanghaijun,Hongming_PhysRevX.5.011029}. A variety of topological materials have been observed or confirmed by modern experimental techniques in the past two decades after theoretical and numerical predictions \cite{qi_RevModPhys.83.1057,kane_RevModPhys.82.3045,molenkamp,hasan_ti,hasan_weyl,dinghong,qah}. However, the above great success still relies on the non-interacting or weak-interacting description of topological insulators. The destiny of topological phases under correlation is now becoming one intriguing question in the topological quantum matter \cite{Rachel_2018}.
Generally speaking, the strongly correlated system is dominated by Mott physics. If we ignore the exotic topological order or quantum spin liquid, Mott physics is close to atomic physics owing to strong electron-electron repulsion \cite{Wen_RevModPhys.89.041004,Zhou_RevModPhys.89.025003,book1}. Based on its definition, a nontrivial topological insulator should be topologically distinct from atomic insulators.
Hence, a topological insulator will close its gap into another Mott phase at large correlation, as illustrated in Fig.\ref{fig0}(a).
A typical example of this is the Kane-Mele-Hubbard model, which ends in an antiferromagnetic Mott insulator at large $U$ \cite{Assaad_PhysRevB.85.115132, Wucj_PhysRevB.84.205121, Kun_PhysRevLett.120.157205, Rachel_2018}.

On the other hand, topological quantum chemistry provides a new insight into our understanding of topological matters \cite{po_nc,Topological_quantum_chemistry,cano_PhysRevB.97.035139}. The basic idea of this approach is constructing the atomic limit of all 230 crystal symmetry groups and exhausting all topological trivial phases. During this process, they proposed there are two different atomic limits: atomic insulator (AI) and
obstruct atomic insulator (OAI) \cite{Topological_quantum_chemistry,cano_PhysRevB.105.125115,xuyuanfeng,Zhijun_Wang}. For the AI limit, the band Wannier centers lie exactly on the atomic sites while they deviate away from atomic sites in the obstruct atomic limit \cite{Topological_quantum_chemistry}. Although OAI does not belong to TI now, it still contains nontrivial boundary states.   Does OAI have different destiny under correlation compared to TI?  The answer to this question is profound and has been believed to ``Yes". Namely , OAI can evolve into the Mott phase without phase transition, as illustrated in Fig.\ref{fig0}(b). Here  we develop a systematic way of 
topological analysis based on many-body wavefunctions  to prove that OAI can smoothly connect to its Mott phase without Mott transition.

\begin{figure}
	\begin{center}
		\fig{3.4in}{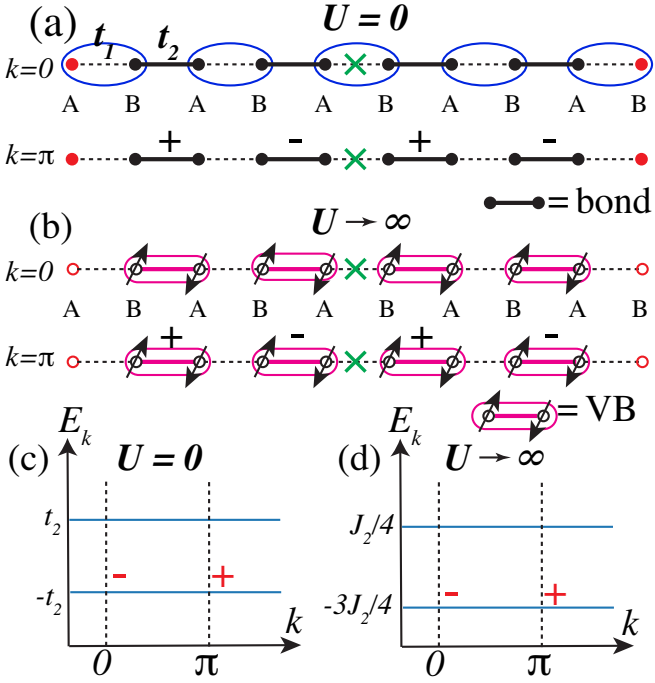}\caption{(a) The noninteracting SSH chain with intra-cell hopping $t_1$ and inter-cell hopping $t_2$. Each unit cell contains two sublattices A and B (inside each blue circle). The red sites are its boundary states. The lower panel is the $k=\pi$ eigenstate with its real space phase factor $e^{ikr_i}$. (b) The valence bond ground state of SSH chain at $t_1=0$ and large U. The red circles are its charge-free boundary states.
  The lower panel is also the $k=\pi$ eigenstate with phase factors defined in Eqn.\ref{vbsk}.
  (c) Band structure $E_k$ of SSH chain at $t_1=0$ and inversion symmetry $\hat{\mathcal{I}}$ eigenvalues at $k=0$, $k=\pi$. Each band is double degenerate for spin. The inversion centers are green crosses in (a-b).
  (d) SSH-Hubbard chain many-body eigenvalues $E_k$ at $t_1=0$ and inversion symmetry $\hat{\mathcal{I}}$ eigenvalues at $k=0$, $k=\pi$. $-\frac{3}{4}J_2$  is single degenerate and $\frac{1}{4}J_2$ is three-fold degenerate.
			\label{fig1}}
	\end{center}
	\vskip-0.5cm
\end{figure}

\textit{1D OAI model}
The prototypical example of OAI is the one-dimensional (1D) Su–Schrieffer–Heeger (SSH) chain, as shown in Fig.\ref{fig1} (a) \cite{ssh, Rice_Mele}. For the SSH chain, if the intra-cell hopping $t_1$ is less than the inter-cell hopping $t_2$, it lies in the OAI phase with boundary states at its ends. 
From the modern theory of polarization, we know that the band Wannier center is directly related to its Berry phase \cite{Resta_RevModPhys.66.899, hughes_PhysRevB.96.245115}. Hence, the Wannier center of OAI occupied band is found to be $\frac{1}{2}$ rather than $0$ in the AI limit \cite{hughes_PhysRevB.96.245115}, as illustrated in Fig.\ref{fig0} (b) inset.

For the interaction SSH chain, we consider the SSH-Hubbard model, 
\begin{eqnarray}
    H_{SSH}=\sum_{i} (t_1 C_{iA, \sigma}^\dagger C_{iB,\sigma}+t_2 C_{iB,\sigma}^\dagger C_{i+1A,\sigma}+h.c.) +U \hat{n}_{i\tau\uparrow}\hat{n}_{i\tau\downarrow}
\end{eqnarray}
where $\sigma$ is the spin index, $i$ is the unit cell index, $\tau=A/B$ is the sublattice index and Einstein summation notation is used here. Various numerical methods have already been applied and found the boundary states at large $U$ limit \cite{Shen_PhysRevB.84.195107,Wcj_PhysRevB.91.115118,Victor_PhysRevB.86.205119,Fan_PhysRevB.94.165167}. Actually, this model contains one exactly solvable limit when $t_1=0$. At this limit, the SSH-Hubbard becomes one decoupled two sites Hubbard problem \cite{book1,book2}. The ground state in this limit is the valence-bond solid (VBS) at large $U$, whose wavefunction can be written as
\begin{eqnarray}
  |VBS\rangle=\prod_{i}(iB,i+1A)
\end{eqnarray}
Here we use the $(a,b)=\frac{|\uparrow_{a}\downarrow_{b}\rangle-|\downarrow_{a}\uparrow_{b}\rangle}{\sqrt{2}}$ symbol for the valence-bond singlet between site $a$ and $b$. Although the charge degree of freedom is gapped, there are still charge-free boundary states at VBS ends, as illustrated in Fig.\ref{fig1}(b).
Hence, the obstructed property remains unchanged at large $U$ limit.

Since the non-interacting band picture is not available here, how to characterize this obstruction at large $U$ limit becomes the essential part.  Entanglement entropy, Green's functions have been applied to this non-trivial property \cite{Wcj_PhysRevB.91.115118,Victor_PhysRevB.86.205119,Fan_PhysRevB.94.165167,Wangzhong_PhysRevX.2.031008}. But these approaches are difficult in linking to the Wannier center and obstruction directly, especially in high dimensions.
On the other hand, symmetry indicators play an important role in identifying topological properties and Wannier centers during the development of topological physics \cite{Liangfu_PhysRevB.76.045302,Topological_quantum_chemistry,po_nc,hughes_PhysRevB.96.245115,fangchen,xiangang_wan,bernevig}.
Embedding symmetry indicators into many-body physics becomes a more suitable way.

Our  key observation here is that translation operator $\hat{T}$ and inversion symmetry $\hat{\mathcal{I}}$ remain good symmetries in the VBS phase.
Using the Bloch theorem, we can construct the momentum eigenstate of VBS as
\begin{eqnarray}
  |VBS\rangle_k=\prod_{i}(iB,i+1A) e^{ikr_i}
  \label{vbsk}
\end{eqnarray}
where $r_i$ is the coordinate of each VB.
This state is the eigenstate of $\hat{T}$ with eigenvalue $e^{ika}$, where $a$ is the lattice constant. One can prove that it remains the ground state of $H_{SSH}$ with eigenvalue $-\frac{3}{4}J_2=-\frac{3t_2^2}{U}$. Using this approach, we can formally plot the many-body eigenvalues $E_k$ as flat bands, as shown in Fig.\ref{fig1}(d). Notice that, $-\frac{3}{4}J_2$ eigenvalue is non-degenerate for spin-singlet while $\frac{1}{4}J_2$ eigenvalue is three-fold degenerate for spin-triplet. This is different from the double degenerate non-interacting band structures in Fig.\ref{fig1}(c).
Using this many-body wavefunction, the $\hat{\mathcal{I}}$ symmetry eigenvalues for $|VBS\rangle_k$ can be found at $k=0$ and $k=\pi$ from its real space pattern in Fig.\ref{fig1}(b). 
Analogy to noninteracting case \cite{hughes_PhysRevB.96.245115}, the obstruction indicator $\nu$ can defined as
\begin{eqnarray}
    e^{i2\pi \nu}=\hat{\mathcal{I}}(0)\hat{\mathcal{I}}(\pi)
\end{eqnarray}
The VBS ground state $\nu$  is $\frac{1}{2}$ as before, which relates to its charge-free boundary states shown in Fig.\ref{fig1}(b).

\begin{figure}
	\begin{center}
		\fig{3.4in}{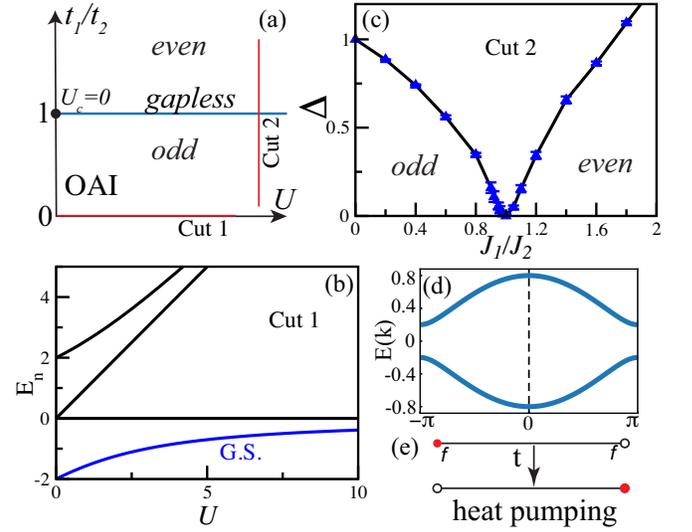}\caption{(a) Phase diagram of SSH-Hubbard model. There are two insulating phases, the ``odd" phase for OAI and the ``even" phase for AI. Their phase boundary is at $t_1=t_2$ blue line with gapless spin excitations.
			(b) Eignestates $E_n$ evolution under $U$ along the Cut 1 in (a). The blue line is the G.S. energy.
			(c) Many-body gap of SSH-spin chain as a function of $J_1/J_2$ along the Cut 2 in (a). 
			(d) The band-structure for the Jordan–Wigner spinless fermions with $J_1=0.6$, $J_2=1$.
                (e) The quantized heat pumping for spinless fermions.
			\label{fig2}}
	\end{center}
	\vskip-0.5cm
\end{figure}

Beyond this exact solvable point, we can obtain the SSH-Hubbard phase diagram using numerical methods \cite{quspin,vmc}. The main results are summarized in Fig.\ref{fig2}(a). There are two insulating phases, the ``odd" phase for OAI and the ``even" phase for AI. 
We first analyze the phase diagram along the Cut 1 line where $t_1=0$. In Fig. \ref{fig2}(b), we plot the ground state (G.S.) energy and three excited state energies as a function of $U$. The G.S. energy is always separated from the first excited state. Therefore, the OAI is smoothly connected with the VBS state without a gap closing, which is consistent with our conjecture in Fig.\ref{fig0}(b). 

Then, we can go through the Cut 2 line at large $U$ limit. In this case, we can map the Hubbard model to the SSH-spin chain model as
\begin{eqnarray}
    H_{J}=\sum_{i} J_1 \mathbf{S}_{iA} \mathbf{S}_{iB}+J_2 \mathbf{S}_{iB} \mathbf{S}_{i+1A}
\end{eqnarray}
where $J_{1/2}=\frac{4t_{1/2}^2}{U}$.
Using the exact diagonalization (ED) \cite{quspin}, we calculated the G.S. state energy (S=0) and first excited state energy (S=1) up to L=32 sites. The gap energy $\Delta$  can be extrapolated by finite size scaling $1/L$ \cite{sandvik}, shown in the supplementary materials (SMs). From Fig.\ref{fig2}(c), the $\Delta$ is only close at $J_1=J_2$ point, where the gapless spin-wave dominates the low-energy excitations \cite{spin-wave_PhysRev.128.2131,nagaosa1999quantum}. Additionally, the solvable limit becomes at $J_1=0$, where the spin-triplet and the spin-singlet gap is $J_2$ as indicated in Fig.\ref{fig1}(d).
Therefore, we can conclude that the phase transition between the odd and even phases is also at $J_1=J_2$, where $t_1=t_2$ is also the non-interacting phase transition point. Since the 1D homologous Hubbard model is exactly solved with Mott transition at $U_c=0$ \cite{Lieb_PhysRevLett.20.1445}, we can finish the phase diagram of Fig.\ref{fig2}(a). The phase transition line between even and odd phases is along the $t_1=t_2$ with gapless spin excitations. From this phase diagram, we can conclude that OAI is smoothly connected with its Mott phase. The atomic obstruction property persists along this evolution. But what is missing in this process?

To answer this question, we can apply the Jordan-Wigner transformation to $H_{J}$ arriving at a spinless fermionic model
\begin{eqnarray}
    H_{f}&=&\sum_{i,\sigma} (\frac{J_1}{2} f_{iA}^\dagger f_{iB}+\frac{J_2}{2} f_{iB}^\dagger f_{i+1A,\sigma}+h.c.)  \nonumber \\ 
    &+&J_1 (n_{iA}-\frac{1}{2})(n_{iB}-\frac{1}{2})+J_2 (n_{iB}-\frac{1}{2})(n_{i+1A}-\frac{1}{2})
\end{eqnarray}
where $f$ is the spinless fermion operator mapping the spin operator like $S_j^{\dagger}=f_j^{\dagger} e^{i \pi \sum_{l<j} n_l}$.
Since there is no magnetic order, we can safely do the mean-field approximation and drop the second density-density term.
The mean-field Hamiltonian is just the SSH model by replace $t_{1/2}$ with $\frac{J_{1/2}}{2}$. We also calculate the Wannier centers using the Wilson loop method \cite{hughes_PhysRevB.96.245115,W_loop_PhysRevB.84.075119} and obtain the same result as the noninteracting case.
It is also widely known that a quantized charge pumping occurs through adiabatic deformations of the SSH model \cite{hughes_PhysRevB.96.245115}. This quantization is related to 2D Hall effect by treating the adiabatic parameter time as the second dimension.
Since the electronic charge is already gapped here, the quantized pumping carried by $f$ becomes the heat as the thermal Hall effect in spin liquid \cite{Subir_PhysRevResearch.2.033283,Zhou_RevModPhys.89.025003}. Hence, the heat pumping is quantized into $\frac{\kappa_{xt}}{T}=-\frac{\pi k_B^2}{6\hbar}$, as illustrated in Fig.\ref{fig2}(e).

\begin{figure}
	\begin{center}
		\fig{3.4in}{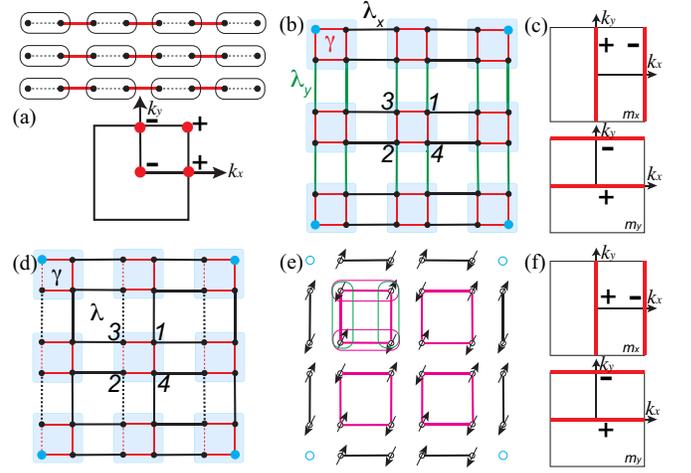}\caption{(a) The stacked SSH-Hubbard chain and its corresponding inversion eigenvalues $\hat{\mathcal{I}}(\mathbf{k})$ in 2D BZ.
    (b) The anisotropic square lattice model with intracell hopping $\gamma$, intercell hopping $\lambda_x$ along $x$ and $\lambda_y$ along $y$. There are four 
 corner states in its odd phases.
    (c) The VB wavefunction mirror symmetry eigenvalues $m_x$ along $k_x=0/\pi$ and $m_y$ along $k_y=0/\pi$ for (b). (d) The quadrupole insulator model with 
    intracell hopping $\gamma$ and intercell hopping $\lambda$. The solid bonds mean hopping sign $>0$ and dash bonds for hopping sign $<0$. There are four 
 corner states as in (b). (e) The plaquette VB G.S. at $\gamma=0$ limit. There are four charge-free corner states at open boundary conditions. The plaquette VBs are linked with pink bonds. The VBs along the edges are linked with black bonds.
    (e) The plaquette VB G.S. mirror symmetry eigenvalues $m_x$ along $k_x=0/\pi$ and $m_y$ along $k_y=0/\pi$.
	\label{fig3}}
	\end{center}
	\vskip-0.5cm
\end{figure}

\textit{2D OAI model}
After finishing the 1D example, we want to generalize our conjecture to 2D. A straightforward generalization is stacking the SSH chain as in Fig.\ref{fig3}(a).
This stacked insulator is basically the 1D physics \cite{hughes_PhysRevB.96.245115}. The similar many-body wavefunctions with translation eigenvalue $e^{i \mathbf{k}\mathbf{a}}$ can be constructed as above, where $\mathbf{k/a}$ are  vectors for momentum and lattice constants in 2D respectively.
The obstruction index is extended by the inversion eigenvalues in the 2D Brillouin zone (BZ) at the high-symmetry points $\Gamma$, $X$, $Y$, and $M$ in Fig.\ref{fig3}(a), which remains same at the VBS phases in Fig.\ref{fig1}(d). Hence, the OAI phase of this stacked 1D chain also connects with the Mott phase with chain boundary states at its edges.

Another interesting OAI model in 2D is the anisotropic square lattice model in Fig.\ref{fig3}(b). Inside each unit cell, there are four sublattices $1-4$ with intracell coupling $\gamma$. The intercell couplings along the $x$ and $y$ directions are anisotropic with $\lambda_x$ and $\lambda_y$. 
This non-interaction model contains non-trivial corner states when $\lambda_x\neq\lambda_y$ and $\gamma<|\lambda_x-\lambda_y|$, as shown in Fig.\ref{fig3}(b).
In this case, there is also one solvable limit at $\gamma=0$ with Hubbard interaction. The ground state here is another 4-sites coupled VBS obtained from ED. The 2D obstruction index is calculated from the mirror symmetry eigenvalues $m_{x/y}$,
\begin{eqnarray}
    e^{i2\pi v_x}=m_x(0,k_y)m_x(\pi,k_y)^*  \nonumber \\
    e^{i2\pi v_y}=m_y(0,k_x)m_y(\pi,k_x)^*  
\end{eqnarray}
Here is one fundamental difference between non-interacting and large $U$ phases. For $U=0$, there are two occupied bands. Each of them has its Wannier centers. At strong correlation, the non-interacting band picture is invalid. All the occupied bands collapse into the nondegenerate ground state and its many-body wavefunction. The $m_{x/y}$ of the anisotropic square lattice model G.S. are plotted in Fig.\ref{fig3}(c). The obstruction index is obtained as $(\frac{1}{2},\frac{1}{2})$ with corner states.

Finally, we consider the quadrupole insulator model in Fig.\ref{fig3}(d) \cite{hughes_PhysRevB.96.245115,hughes_science, Ezawa, c3}. Compared to the anisotropic square lattice model, the hopping signs along the $y$ direction are flipped along the dash bonds and take $\lambda_x=\lambda_y=\lambda$. The quadrupole 4-band Bloch Hamiltonian can be written as
\begin{eqnarray}
    H(\mathbf{k})_{qp}&=&[\gamma+\lambda \cos k_x] \Gamma_4+\lambda \sin k_x \Gamma_3 \nonumber \\
    &+&[\gamma+\lambda \cos k_y] \Gamma_2+\lambda \sin k_y \Gamma_1
\end{eqnarray}
where $\Gamma_l=-\tau_2\sigma_l$ for $l=1\sim3$ and $\Gamma_4=\tau_1\sigma_0$. $\tau$, $\sigma$ are Pauli matrices for the degrees of freedom within a unit cell. $H_{qp}$ is invriant respect to two mirror symmetries $M_x=\tau_1\sigma_3$ and $M_y=\tau_1\sigma_1$.

This model hosts the corner states when $\gamma<\lambda$, as illustrated in Fig.\ref{fig3}(d). Adding the Hubbard interaction, we can also approach the OAI from the $\gamma=0$ limit. The ground state at large $U$ becomes the plaquette VB solid, as shown in  Fig.\ref{fig3}(e). The wavefunction for each plaquette can be written as
\begin{eqnarray}
    |pVB\rangle=(1,4)(3,2)-(1,3)(4,2)
\end{eqnarray}
The G.S. wavefunction is also the product state $\prod_{i}|pVB\rangle_i$. The corresponding mirror symmetry $m_{x/y}$ values are plotted in Fig.\ref{fig3}(f).
Hence, the obstruction index remains the same. The corner states now become the charge-free corner states with open boundaries, as illustrated in Fig.\ref{fig3}(e). Along its four edges, the bonds become 1D VB as in the SSH-Hubbard besides the plaquette VB inside bulk.

\begin{figure}
	\begin{center}
		\fig{3.4in}{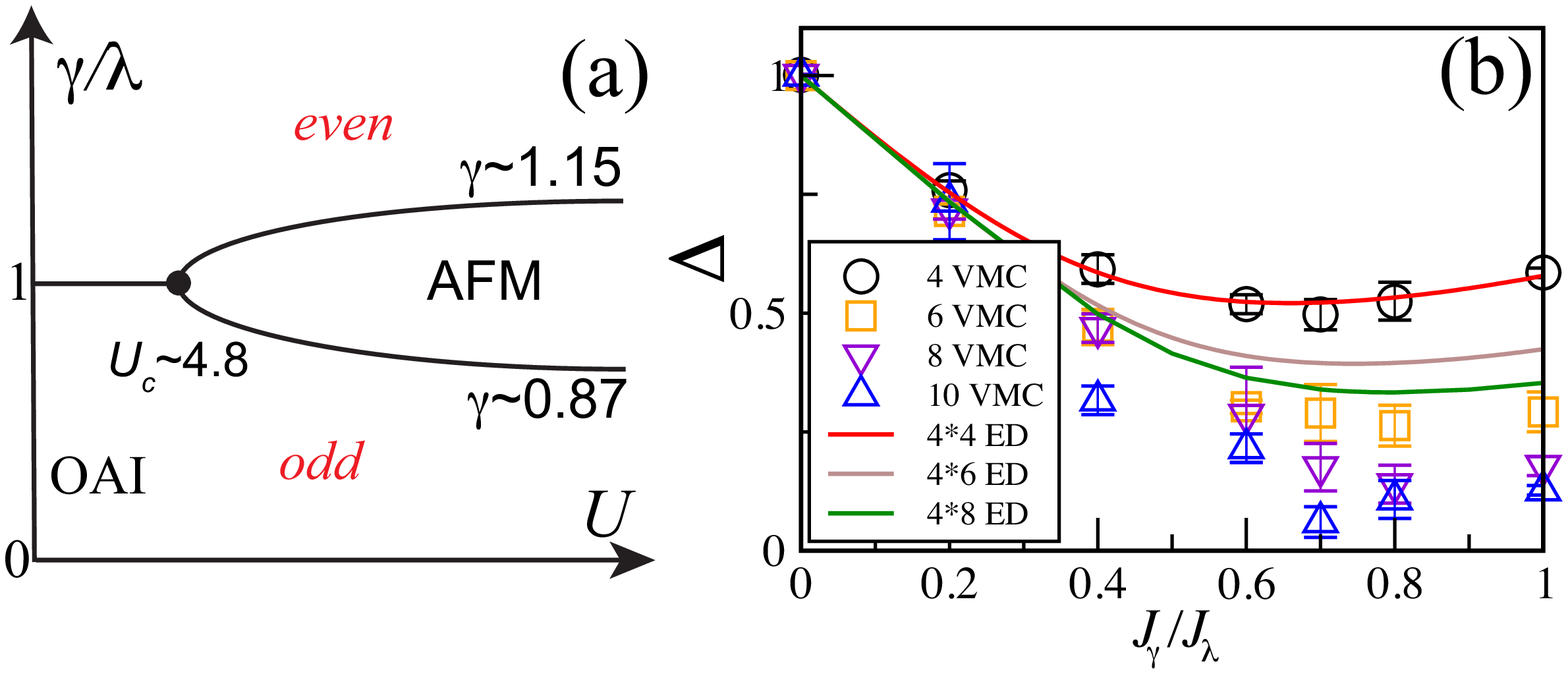}\caption{(a), The phase diagram of quadrupole-Hubbard model with lattice in Fig.\ref{fig3}(d). At small $U$, the ``even" and ``odd" phases are separated by $\gamma=\lambda$ line with unit $\lambda=1$. At large $U$, AFM states split the phase diagram into two phase transition lines approaching $\gamma\sim0.87$ and $\gamma\sim1.15$. (b) The spin gaps $\Delta$ of $J_{\lambda/\gamma}$ Heisenberg model obtained from $4\times L$ ED and $L\times L$ VMC. $\Delta$ is supposed to vanish in AFM and finite in VBS. The turning point is around $0.75 \pm 0.05$.
			\label{fig5}}
	\end{center}
	\vskip-0.5cm
\end{figure}

The phase diagram of this quadrupole-Hubbard model is sketched in Fig.\ref{fig5}(a). The line along $\gamma=0$ without phase transition is calculated in SM. This OAI phase also evolves into the Mott phase without phase transition using ED, as we claimed in Fig.\ref{fig0}.
At the large $U$ limit, the Hubbard model maps into the Heisenberg model with couplings $J_{\lambda/\gamma}=\frac{4(\lambda/\gamma)^2}{U}$ along $\lambda/\gamma$ bonds. Both ED and variational Monte Carlo (VMC) methods \cite{quspin,vmc} have been applied to this Heisenberg model. The spin gaps $\Delta$ at finite-size lattices are plotted in Fig.\ref{fig5}(b).
For the 2D square lattice Heisenberg model, the G.S. is the antiferromagnetic (AFM), where the spin gap vanishes owing to the spin wave.
The $\Delta$ in Fig.\ref{fig5}(b) has turning points at finite-size lattices around $J_\gamma\sim(0.75 \pm 0.05) J_\lambda$. Since VMC is one approximate method, the transition value has a large variance. The stable AFM region is still reasonable compared to 1D SSH.
In 1D, the Mermin–Wagner theorem tells us that the long-range gapless fluctuations kill the magnetic order. Hence, the gapped VBS is more stable with transition pined to $J_1=J_2$. In 2D, the G.S. AFM is stable at zero temperature  with weaker gapless fluctuations. Hence, AFM could extend finite phase space. 

Therefore, there are two phase transition lines at large $U$ with $\gamma\sim0.87$ and $\gamma\sim1.15$ from the Heisenberg model results.
Another phase transition line is along $\gamma=\lambda$ at finite $U$. The non-interacting band structures close gap with two Dirac cones at $(\pi,\pi)$. The Mott transition here is similar to the graphene-Hubbard model with finite $U_c$ owing to zero density of states \cite{Sen_PhysRevB.89.195119,Assaad_PhysRevB.85.115132}. The $U_c$ is found around $4.8$ using slave-boson mean field calculation. After this point, the phase line splits into two transition lines approaching the large $U$ values as shown in Fig.\ref{fig5}(a). From this phase diagram, we can conclude that OAI still smoothly evolves into the Mott phase in a wide-range phase diagram. But it is still possible towards a gapless magnetic order with gap closing \cite{Luzhongyi_PhysRevB.102.045110}. 

The above conclusions can be further extended to 3D. The octupole model with corner states in 3D is also studied in SM. The phase diagram is sketched in a similar pattern as Fig.\ref{fig5}(a) with the AFM phase. The large $U$ limit with $\gamma=0$ is also exactly solvable. The G.S. can be calculated using ED. The G.S. wavefunction can be approximated by the dimer-RVB state inside the cubic \cite{Delgado_PhysRevB.56.8774}. The symmetry eigenvalues can also be found accordingly.

In summary, we study the correlated obstructed atomic insulators. Contradicted to topological insulators without atomic limits, 
the OAI can smoothly connect with Mott phases without phase transition.
The obstruction properties with boundary modes persist at large $U$ when the charge freedom is gapped. The symmetry indicators through the many-body wavefunctions in momentum space are used to prove the obstruction properties. The SSH, the anisotropic square lattice model, the quadrupole insulator model, etc. 
have been studied with exact diagonalization and variational Monte Carlo. 
We hope that these findings could provide a new understanding of correlated atomic insulators.

\textit{Acknowledgement}
This work is supported by the Ministry of Science and Technology  (Grant No. 2022YFA1403901, No.2022YFA1403800), the National Natural Science Foundation of China (Grant No. NSFC-11888101, No. NSFC-12174428), the Strategic Priority Research Program of the Chinese Academy of Sciences (Grant No. XDB28000000, XDB33000000), the New Cornerstone Investigator Program, and the Chinese Academy of Sciences through the Youth Innovation Promotion Association (Grant No. 2022YSBR-048).

\bibliographystyle{apsrev4-1}
\bibliography{reference}

\clearpage
\onecolumngrid
\begin{center}
\textbf{\large Supplemental Material: The destiny of obstructed atomic insulator under correlation}
\end{center}

\setcounter{equation}{0}
\setcounter{figure}{0}
\setcounter{table}{0}
\setcounter{page}{1}
\makeatletter
\renewcommand{\theequation}{S\arabic{equation}}
\renewcommand{\thefigure}{S\arabic{figure}}
\renewcommand{\bibnumfmt}[1]{[S#1]}
\renewcommand{\citenumfont}[1]{S#1}

\twocolumngrid
\section{ED for spin gap in 1D chain}
\begin{figure}
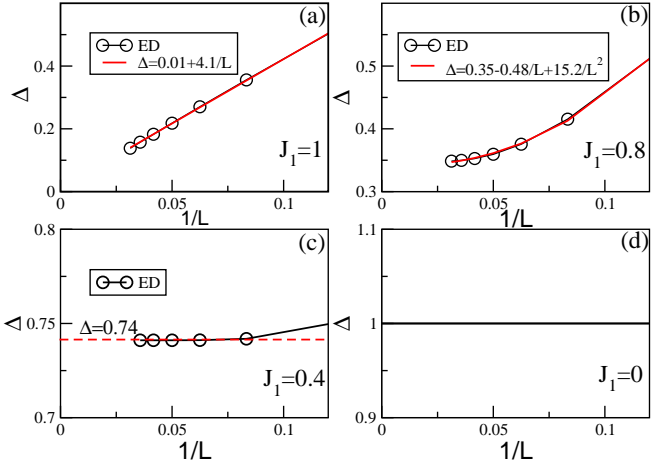

	\begin{center}
		\fig{3.4in}{spinchain.eps}\caption{ The spin gaps at various $J_1$ as function of $1/L$ and their fitting functions from ED.
   \label{fig_chain}}
	\end{center}
	\vskip-0.5cm
\end{figure}

In this section, we plot the spin gap $\Delta$ for the L site SSH-spin chain model $H_J$ with periodic boundary in Fig.\ref{fig_chain}. The spin gap is between $S=0$ ground state and the first excited states $S=1$. It is also possible the first excited states in $S=0$. But it is always degenerate or higher than $S=1$ from our ED calculations.
Then finite-size scaling functions are used to fit the spin gap $\Delta$ as a function of $1/L$. For $J_1=1$, the model goes back to the AFM spin chain with gapless excitations. The linear fit in Fig.\ref{fig_chain}(a) leads to a gap on the order of 0.01, which is consistent with the gapless feature. The $J_1=0.8$ fitting function becomes a power law with a gap around 0.35 in Fig.\ref{fig_chain}(b). The gap in Fig.\ref{fig_chain}(c) with $J_1=0.4$ is saturated in 0.74. The gap at $J_1=0.0$ is exactly solvable with $\Delta=1$. Using this method, we obtain the gap values Fig.\ref{fig2}(c) in the main text.

\section{Hamiltonian for square lattice and octupole model}
The Bloch Hamiltonian of the anisotropic square lattice model is
\begin{eqnarray}
    H_{sq}(\mathbf{k})&=&\left(\begin{array}{cc} 
		0 & q(\mathbf{k})\\
		q(\mathbf{k})^\dagger & 0 
	\end{array}\right)  \nonumber\\
    q(\mathbf{k})&=&\left(\begin{array}{cc} 
		\gamma+\lambda_x e^{ik_x} & \gamma+\lambda_y e^{ik_y}\\
		\gamma+\lambda_y e^{-ik_y} & \gamma+\lambda_x e^{-ik_x} 
	\end{array}\right) 
\end{eqnarray}

\section{The octupole model}
\begin{figure}
	\begin{center}
		\fig{3.4in}{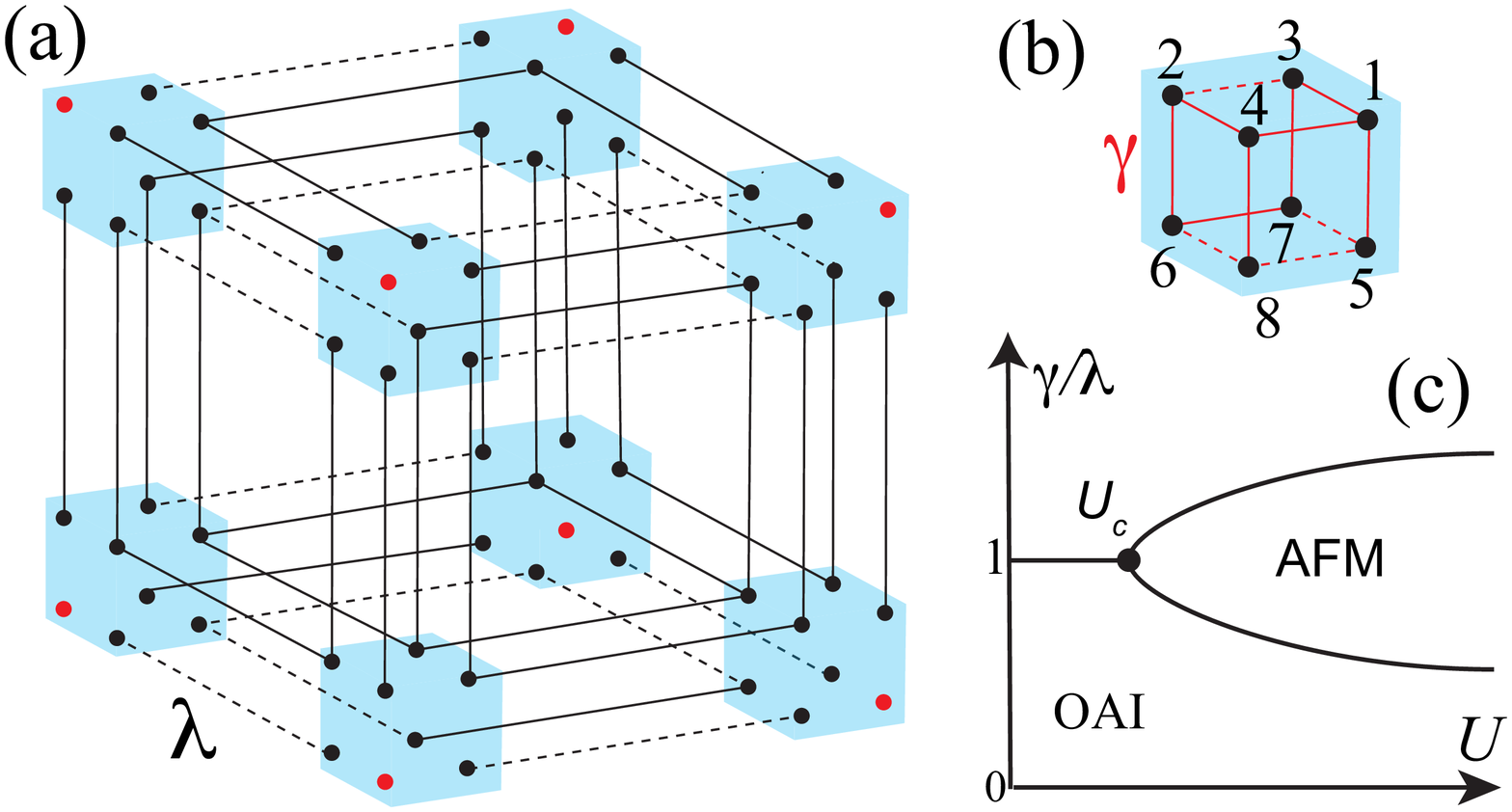}\caption{(a) The lattice of octupole model. (b) The eight sites inside the octupole model (a) with intra-cell coupling $\gamma$
(c) The sketched phase diagram of the octupole-Hubbard model		
   \label{figoct}}
	\end{center}
	\vskip-0.5cm
\end{figure}

The lattice octupole model is shown in Fig. \ref{figoct}(a-b). The Bloch Hamiltonian of the octupole model is
\begin{eqnarray}
    H_{oct}=\lambda_y \sin(k_y) \Gamma^{'1}+[\gamma_y+\lambda_y \cos(k_y)]\Gamma^{'2}  \\
    +\lambda_x \sin(k_x) \Gamma^{'3}+[\gamma_x+\lambda_x \cos(k_x)]\Gamma^{'4} \\
    +\lambda_z \sin(k_z) \Gamma^{'5}+[\gamma_z+\lambda_z \cos(k_z)]\Gamma^{'6}
\end{eqnarray}
where $\Gamma^{'i}=\sigma_3\otimes\Gamma_i$ for $i=0,1,2,3$, $\Gamma^{'4}=\sigma_1 \otimes I_{4\times4}$, $\Gamma^{'5}=\sigma_2 \otimes I_{4\times4}$, and
$\Gamma^{'6}=i\Gamma^{'0}\Gamma^{'1}\Gamma^{'2}\Gamma^{'3}\Gamma^{'4}\Gamma^{'5}$. We can choose all $\lambda$ and $\gamma$ equal. The OAI phase with corner states lies in 
$\lambda>\gamma$.
The octupole-Hubbard model phase diagram is sketched in Fig.\ref{figoct}(c), with a similar phase diagram of quadrupole-Hubbard. The exact solvable limit lies in $\gamma=0$ as discussed below.

\begin{figure}
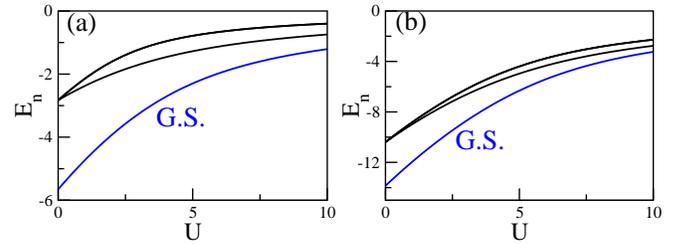

	\begin{center}
		\fig{3.4in}{E48.eps}\caption{(a) Energies of the quadrupole-Hubbard model at $\gamma=0$ through exact diagonalization. (b)  Energies of the octupole-Hubbard model at $\gamma=0$ through exact diagonalization. The G.S. eigenvalues are highlighted using blue lines.
			\label{figED}}
	\end{center}
	\vskip-0.5cm
\end{figure}

\section{Eigenvalues for the quadrupole Hubbard and octupole Hubbard model}
The eigenvalues for the quadrupole Hubbard and octupole Hubbard models with $\gamma=0$, obtained from ED, are shown in Fig.\ref{figED}. 
In this limit, the models become a $L=4$ periodic spin chain (4 sites) and a $L=4$ periodic spin ladder (8 sites).
There are no phase transitions under correlation as we claimed in the main text.

\section{numerical detials}
The ED method we used here is implemented using the quspin package \cite{quspin}, where the symmetries have already been applied \cite{sandvik}.

The VMC method used here is implemented using the mVMC package \cite{vmc}. The mVMC can simultaneously optimize  many variational variables and find the variational wavefunctions we want.

\section{$C_3$ obstructed atomic insulator}
\begin{figure}
	\begin{center}
		\fig{3.4in}{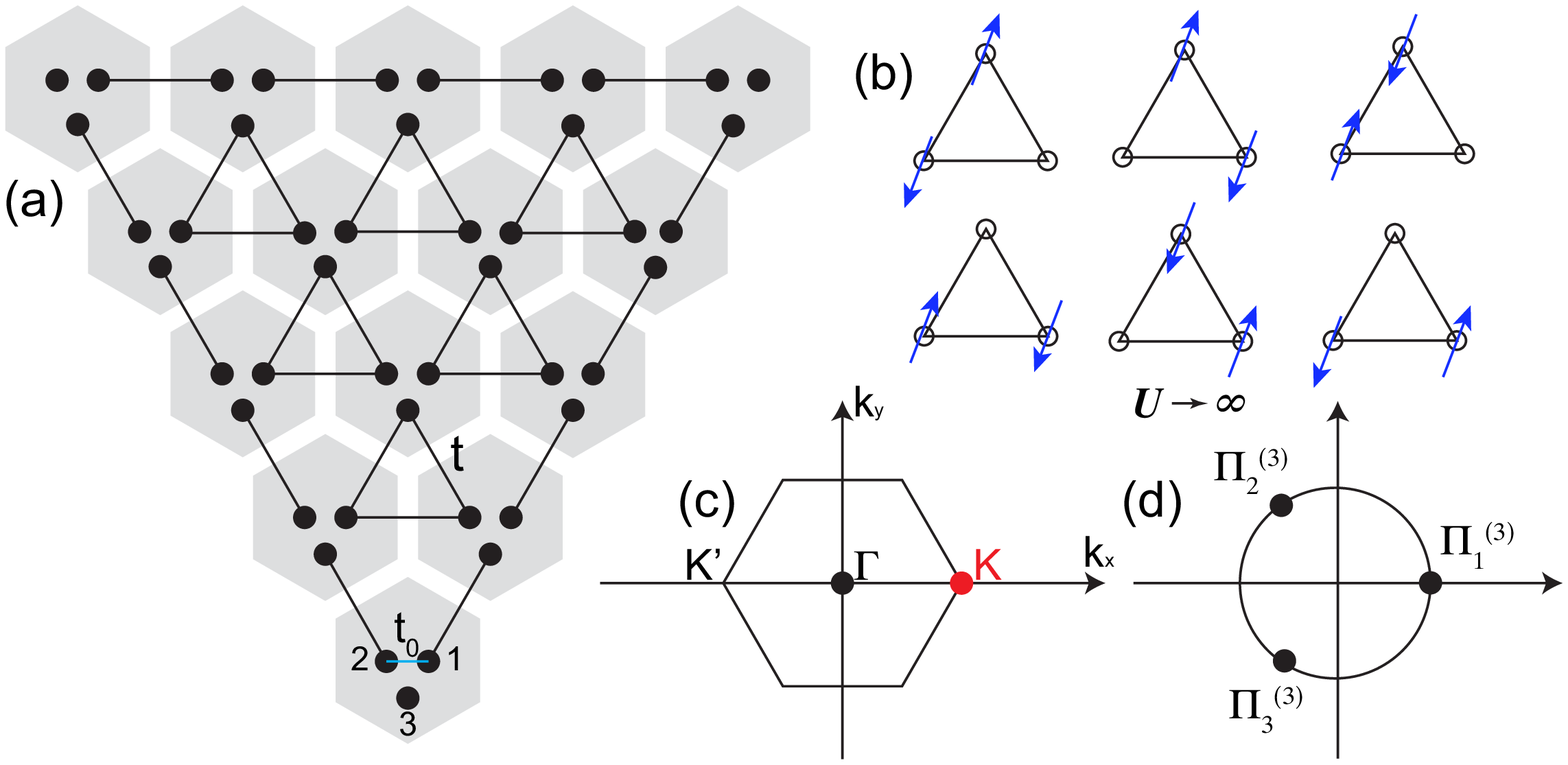}\caption{(a) $C_3$  OAI with inter-cell coupling $t$ and intra-cell coupling $t_0$. (b)  The ground state wavefucntion at large U and $t_0=0$, $t=-1$ is the linear combination of 6 configurations.
			(c) The BZ of $C_3$ OAI. (d)  $C_3$  Rotation eigenvalues at high symmetry points.
			\label{figc3}}
	\end{center}
	\vskip-0.5cm
\end{figure}

In this section, we discuss the $C_3$ OAI \cite{Ezawa,c3} as shown in Fig. \ref{figc3} (a). The inter-cell coupling is $t=-1$ and the intra-cell coupling is $t_0$. The Bloch Hamiltonian of this model is
\begin{eqnarray}
	H_{c3}(\mathbf{k})&=&\left(\begin{array}{ccc} 
		0 & t_0+t e^{i \mathbf{k}\cdot \mathbf{a}_1} &  t_0+t e^{i \mathbf{k}\cdot \mathbf{a}_2}  \\
		 t_0+t e^{-i \mathbf{k}\cdot \mathbf{a}_1}  & 0 &  t_0+t e^{-i \mathbf{k}\cdot \mathbf{a}_3} \\
		 t_0+t e^{-i \mathbf{k}\cdot \mathbf{a}_2} & t_0+t e^{i \mathbf{k}\cdot \mathbf{a}_3} & 0
	\end{array}\right) 
\end{eqnarray}
where $\mathbf{a}_1=(1,0)$, $\mathbf{a}_2=(\frac{1}{2},\frac{\sqrt{3}}{2})$, $\mathbf{a}_3=\mathbf{a}_1-\mathbf{a}_2$.
 This $C_3$ model lies in the OAI phase when $|t_0|<1$ with corner states shown in Fig.\ref{figc3}(a).
 In this case, only the lowest band is fully occupied with 2 electron filling. Since this model is not half-filling, we can not apply the Heisenberg model to the large $U$ limit. 
 However, we still can use the ED to find the ground state at large $U$ at $t_0$ limit.
 Interestingly, the ground state wavefunction for the triangle with 2 electron filling is an equal weight linear combination of  electron configurations in Fig. \ref{figc3}(b).
Then, we can also use the symmetry indicators here \cite{c3}. In the $C_3$ OAI, we use the  $C_3$ rotation eigenvalues at high symmetry points $\Pi$ as plotted in Fig. \ref{figc3}(d).
\begin{eqnarray}
	\Pi_p^{(3)}= e^{2\pi i(p-1)/3}
\end{eqnarray}
where  $p=1,2,3$.
The topological invariants are defined for difference with BZ center $\Gamma$ as 
\begin{eqnarray}
	[\Pi_p^{(3)}]=\Pi_p^{(3)}-\Gamma_p^{(3)}
\end{eqnarray}
The obstructed property is obtained from the $K=(\frac{4\pi}{3},0)$ in BZ (Fig. \ref{figc3}(c))
as
\begin{eqnarray}
	\chi^{(3)}=(K_1^{(3)},K_2^{(3)})
\end{eqnarray}
For the solvable limit in Fig.\ref{figc3}(b), we find 	$\chi^{(3)}=(-1,0)$.

\end{document}